\documentclass[aps,prd,nofootinbib,twocolumn,reprint,groupedaddress,showpacs,showkeys,floatfix]{revtex4-1}
\usepackage {amsmath}
\usepackage {amssymb}
\usepackage {amsfonts}
\usepackage {amsthm}
\usepackage {mathrsfs}
\usepackage {natbib}
\usepackage {latexsym}
\usepackage {graphicx}
\usepackage {txfonts}
\usepackage {rotating}
\usepackage {wasysym}
\usepackage {multirow}
\usepackage {hhline}
\usepackage {hyperref}
\usepackage[usenames,dvipsnames]{xcolor}
\usepackage {bm}
\usepackage{appendix}
\usepackage{acronym}
\usepackage{dcolumn}   
\usepackage {url}
\usepackage {caption}
\usepackage {subcaption}
\usepackage {mathtools}
\usepackage {enumerate}
\usepackage {acronym}
\pdfoutput=1
\newcommand{\dcc}{LIGO-P1500247-v1}

\begin{document}

\title{Systematic errors in estimation of gravitational-wave candidate significance}
\author{C.~Capano}
\affiliation{Max Planck Institute for Gravitational Physics,
  Callinstra\ss e 38, D-30167, Hannover, Germany}
\author{T.~Dent}
\affiliation{Max Planck Institute for Gravitational Physics,
  Callinstra\ss e 38, D-30167, Hannover, Germany}
\author{C.~Hanna}
\affiliation{The Pennsylvania State University, University Park, PA 16802, USA}
\author{M.~Hendry}
\affiliation{School of Physics and Astronomy, Kelvin Building,
  University of Glasgow, Glasgow, G12 8QQ, UK}
\author{Y.-M.~Hu}
\email{huyiming@mail.sysu.edu.cn}
\affiliation{TianQin Research Center for Gravitational Physics, Sun Yat-sen University,
  Zhuhai, 519082, China}
\affiliation{School of Physics and Astronomy, Kelvin Building,
  University of Glasgow, Glasgow, G12 8QQ, UK}
\affiliation{Max Planck Institute for Gravitational Physics,
  Callinstra\ss e 38, D-30167, Hannover, Germany}
\author{C.~Messenger}
\affiliation{School of Physics and Astronomy, Kelvin Building,
  University of Glasgow, Glasgow, G12 8QQ, UK}
\author{J.~Veitch} 
\affiliation{School of Physics and Astronomy,
  University of Birmingham, Edgbaston, Birmingham, B15 2TT, UK}
\affiliation{School of Physics and Astronomy, Kelvin Building,
  University of Glasgow, Glasgow, G12 8QQ, UK}
\date{\today\\\mbox{\dcc}}

\begin{abstract}
The statistical significance of a candidate gravitational-wave (GW) event is crucial to the prospects for a confirmed detection, or for its selection as a candidate for follow-up electromagnetic observation. To determine the significance of a GW candidate, a ranking statistic is evaluated and 
compared to an empirically-estimated background distribution, yielding a false alarm probability or $p$-value.  The reliability of this background estimate is limited by the number of background samples and by the fact that GW detectors cannot be shielded from signals, making it impossible to identify a pure background data set.  Different strategies have been proposed: in one method, \textit{all samples}, including potential signals, are included in the background estimation, whereas in another method, \textit{coincidence removal} is performed in order to exclude possible signals from the estimated background. 
Here we report on a mock data challenge, performed prior to the first
detections of GW signals by Advanced LIGO, to compare these two methods. 
The all-samples method is found to be self-consistent in terms of the rate of false positive detection claims, but its $p$-value estimates are systematically conservative and subject to higher variance.  Conversely, the coincidence-removal method yields a mean-unbiased estimate of the $p$-value but sacrifices self-consistency.  We provide a simple formula for the uncertainty in estimate significance and compare it to mock data results.  Finally, we discuss the use of different methods in claiming the detection of GW signals.
\end{abstract}

\maketitle 

\acrodef{GW}[GW]{gravitational wave}
\acrodef{MDC}[MDC]{mock data challenge}
\acrodef{SNR}[SNR]{signal-to-noise ratio}
\acrodef{CBC}[CBC]{compact binary coalescence}
\acrodef{NS}[NS]{neutron-star}
\acrodef{LIGO}[LIGO]{Laser Interferometer Gravitational-wave
  Observatory}
\acrodef{LSC}[LSC]{LIGO Scientific Collaboration}
\acrodef{LVC}[LVC]{LSC (LIGO Scientific Collaboration) and Virgo collaborations}
\acrodef{CDF}[CDF]{cumulative distribution function}
\acrodef{PDF}[PDF]{probability distribution function}
\acrodef{FAR}[FAR]{false alarm rate}
\acrodef{FAP}[FAP]{false alarm probability}
\acrodef{aLIGO}[aLIGO]{Advanced LIGO}
\acrodef{adV}[adV]{Advanced Virgo}
\acrodef{ROC}[ROC]{Receiver Operating Characteristic}
\acrodef{FPR}[FPR]{false positive rate}
\acrodef{TPR}[TPR]{true positive rate}
\acrodef{IQR}[IQR]{inter-quartile range}
\acrodef{APC}[APC]{all possible coincidences}
\acrodef{CMB}[CMB]{cosmic microwave background}
\acrodef{IFO}[IFO]{interferometer}
\acrodef{BNS}[BNS]{binary neutron star}
\acrodef{NSBH}[NSBH]{neutron star black hole}
\acrodef{IFAR}[IFAR]{inverse FAR}

%
%
\section{Introduction\label{sec:introduction}}
%
%
The first direct detections of \ac{GW} signals from the mergers of massive black hole binary systems achieved substantial scientific and societal impact when reported in 
2016~\cite{GW150914_paper,
TheLIGOScientific:2016agk,
TheLIGOScientific:2016qqj,
TheLIGOScientific:2016wfe,
GW150914:TGR,
Abbott:2016nhf,
TheLIGOScientific:2016uux,
TheLIGOScientific:2016zmo,
Abbott:2016jsd,
TheLIGOScientific:2016htt,
GW150914_stoch,
2016PhRvL.116x1103A,
2016arXiv160604856T,
Harry:2010zz,TheLIGOScientific:2014jea,TheVirgo:2014hva,Aasi:2013wya}. 
A key aspect of the data analysis required for these detections, and an essential step in verifying their identity as astrophysical \ac{GW} signals, was robustly establishing the statistical significance of candidate events in the search for \ac{GW} from binary mergers. 

%
%
Extraordinary claims require extraordinary evidence, and 
the discovery of a previously unseen effect or signal has generally required the significance to exceed a certain threshold.
Consequently, in order to reject confidently the null hypothesis that one's dataset contains \textit{only} noise and claim the presence of a signal, the probability of obtaining this dataset when the null hypothesis is true, i.e. the \textit{p-value}, must be no larger than a 
pre-determined threshold.  The threshold for discovery is often set at ``5 sigma'', i.e.\ the probability of obtaining a value 5 or more standard deviations from mean of a Gaussian distribution~\cite{Lyons2013_5Sigma}, as for instance 
in the detection of the Higgs boson~\cite{Aad2012} and B-modes in the polarization of the \ac{CMB}~\cite{Ade2014}.  Less stringent significance thresholds have typically been applied when presenting \textit{evidence} for a possible new effect, and for applications where the rate of false positives is not required to be extremely low -- for example in selecting candidate \ac{GW} events for follow-up observations over the electromagnetic spectrum~\cite{Abbott:2016gcq}. 

Ground-based detector GW data analysis presents several unique challenges that complicate the calculation and interpretation of the significance. Due to the noisy local environments at each detector, the simultaneous observation of candidate events (or \textit{`triggers'}) by multiple detectors, with consistent signal characteristics observed at each detector, is a necessary condition.  Moreover, the low signal-to-noise ratio (SNR) expected means that matched-filtering of the data with signal templates is required, but similarity among the templates used might lead to correlation of the triggers' SNR. Also, no \textit{a priori} information about the background distribution is available, so this has to be estimated from the observations themselves, and the possible contamination by a real GW signal cannot simply be spotted and removed. The expected rate of \ac{GW} signals, prior to any positive detection, is also uncertain by several orders of magnitude~\cite{ratesdoc}. 

To address these issues, analysis methods have been developed to discover potential GW candidates and assign significance to them \cite{Babak:2012zx,Abbott:2009qj,Cannon2013FAR,Farr:2015}. 
These methods use similar ideas, albeit with quite different specific implementations. More importantly, however, the methods encapsulate two different philosophical viewpoints about the correct way to calculate the significance of a candidate event.  In this manuscript we describe a \ac{MDC}, motivated by seeking to resolve the conflict between these two viewpoints, which emerged in the ``Blind Injection'' exercise undertaken by LIGO and Virgo in 2010~\cite{Colaboration:2011np}.  We demonstrate quantitatively the pros and cons of both viewpoints for quantifying the significance of \ac{GW} candidates via a \emph{false alarm probability} or $p$-value. 

This \ac{MDC} was designed, executed and concluded before the first \ac{GW} detections and its results informed the confident detections of GW150914, GW151226 and GW170104, as well as the evaluation of the less significant candidate event LVT151012.  With detections of \ac{BNS} and \ac{NSBH} events also expected in the next few years, our \ac{MDC} will provide the GW community with clear guidelines for computing and interpreting the false alarm probabilities of future GW detection candidates as well as justifying the choices made for previous results.  A detailed account of the setup, results and conclusions from the \ac{MDC} can be found in~\cite{HamletLong}; here we summarize the essential features of the study and describe the main results relevant to \ac{GW} detection. 

\section{The Mock Data Challenge}
%
%
\subsection{Rationale}
In order to assess the significance of a GW trigger, one must understand the statistical properties of the distribution of background noise. Unfortunately one cannot simply shield the detectors from GW signals and thus measure a background that is guaranteed to be `signal-free'.  However, various methods have been developed to approximate such a background \cite{Babak:2012zx,Abbott:2009qj,Cannon2013FAR}. 

Ground-based GW interferometers constantly generate triggers, but only those triggers that excite a high response in more than one detector simultaneously, in a way consistent with an astrophysical signal, are regarded as viable GW candidates.  These are then known as coincident or \textit{zero-lag} triggers.  The controversy around background estimation focuses on how to deal with these zero-lag triggers.

One approach is simply to include zero-lag triggers in the background estimation -- consistent with the argument that no trigger will be classified as a confirmed signal until it passes a 
pre-determined threshold, so no trigger should be removed prior to that confirmation. This approach might result in actual GW signals contaminating the estimation of the noise background, but in principle this should be a \textit{conservative} contamination -- i.e. any trigger identified as passing the 
threshold would also have done so with an uncontaminated background.

The other approach is to use only single-\ac{IFO} triggers to estimate the noise background, having first removed the zero-lag triggers -- arguing that the latter are intrinsically different as they are (the only) viable GW candidates and thus it is inappropriate to consider them when estimating the background distribution.  In principle, removing background-induced zero-lag triggers should result in an unbiased estimate of the significance of a GW candidate.
 
%
%
\subsection{MDC Setup and Definitions}
To better understand, and hopefully resolve, the potential impact of zero-lag trigger removal on the estimation of significance, we therefore sought to carry out a \ac{MDC} in which all significance calculations using the mock data would be done using both `coincidence-removal' and `all-coincidence' methods.  Since our focus was on this one issue, we recognised that an end-to-end simulation of GW detection would involve lots of technical details not directly linked to the specific question of `removal' versus `non-removal'.  Thus we sought to make the \ac{MDC} as simple as possible.

In generating our mock data, each trigger was labeled with two quantities: a ranking statistic (or SNR) $\rho_i$ for the $i^{\rm th}$ \ac{IFO}, and an arrival time.
A background trigger will occur at each \ac{IFO} independently while an astronomical signal would trigger responses dependent on the geometry of the detectors, as well as that of the source. The frequency of background triggers and astronomical signals are each controlled by a rate parameter.
To mimic choices typically used in LIGO-Virgo analysis~\cite{TheLIGOScientific:2016qqj}, we set a threshold of $\rho^{\rm th}_i=5.5$ for a trigger to be registered in a single \ac{IFO}.

%
%
A time window based on the light travel time across the Earth was used to identify all coincident triggers among different detectors. 
The SNR, $\rho$, of the coincident trigger is the root sum square of the SNR, $\rho_i$, for the individual detectors. This ranking statistic $\rho$ for each coincident pair of triggers is used to calculate the significance of the zero-lag events.

The parameters of the \ac{MDC} were chosen so that each \textit{realisation} contained $\sim 10^4$ background triggers in each \ac{IFO}, and statistically $\sim 10$ zero-lag triggers were expected to originate from random coincidences in the background distribution. Each \emph{experiment} then contained $10^5$ realisations.
This choice allowed us to study different interesting regimes, while keeping the computational burden of the \ac{MDC} practically feasible.  

In total we designed $14$ experiments with background distributions of different levels of complexity, which we label respectively `simple', `realistic' and `extreme'; the astronomical signal rate in these experiments ranges from zero (i.e. no astronomical events at all) through low (an expected $0.001-0.1$ coincidences per realisation), medium ($\sim 0.5$ coincidences per realisation) to high ($\sim 3$ coincidences per realisation).  Crucially, the fact that the background distribution was known to the \ac{MDC} designers made it possible to calculate the exact dependence of the \ac{FAP} on the $\rho$ value for each experiment. 

%
%
We injected \ac{BNS} signals with anticipated \ac{SNR} calculated from the inspiral part of the signal only, assuming random sky locations and distances, uniform in comoving volume, up to a cutoff.  The actual measured \ac{SNR} was randomly assigned based on the simulated signal.  The distributions for both signals and noise were taken to be rapidly decreasing at high SNR, but the distribution of astronomical signals had a shallower (power-law) slope: thus signals `stand out' in the region with larger $\rho$. 

Of course some astronomical events might cause a trigger in one detector only, as the \ac{SNR} in one or more other detectors may not be high enough to exceed the threshold.  Since in this MDC we limit our investigation only to the impact of removal or non-removal on the \emph{loudest} (i.e. highest SNR $\rho$) event, which is not expected to originate from such a signal-noise hybrid, we can safely ignore any adverse impact of this scenario.

\subsection{MDC participants}
%
%
We define the significance of the loudest coincidence in terms of its \ac{FAP}. That is, the probability of observing a coincidence with $\rho$ higher than the recorded loudest zero-lag event.

Three different algorithms for estimating the background distribution from the observed triggers were investigated in the \ac{MDC}.
\begin{enumerate}
\item
The `time slide' algorithm adapted from the all-sky LIGO-Virgo search pipeline~\cite{Babak:2012zx,Abbott:2009qj} shifts one detector's triggers by a certain time interval so that, in the shifted data, no new coincidence could be associated with a real astronomical event.  Thus the statistical properties of the background may be estimated. A \ac{FAR} is calculated for a coincidence with a particular value of $\rho$ by evaluating the frequency of coincidences per analysis time that have as high or higher a value of $\rho$. 
The \ac{IFAR} algorithm is used to estimate the probability of obtaining one or more coincidences with higher \ac{SNR}. By exhausting all allowable time shifts, one can estimate both the expected number of background coincidences and the probability for each to exceed a given $\rho$. 
The \ac{FAP} is then calculated by taking account of the number of trials from the expected coincidence number.
\item
The `\ac{APC}' algorithm used a similar strategy to the time slide algorithm, by combining all physically unrelated triggers from different detectors to form an ensemble of all possible coincidences. 
The \ac{FAP} was then calculated simply by counting the fraction of coincidences louder than the loudest observed event, with trial factors included.  In addition to the \ac{FAP}, the \ac{APC} algorithm can also provide an estimate of its associated uncertainty.  The smaller the \ac{FAP} is, the higher the relative uncertainty will be.
In this \ac{MDC}, the \ac{APC} algorithm dynamically adjusted the calculation precision, so that the estimation of larger \ac{FAP} had a higher uncertainty.
\item
A modification of the \textit{gstlal} algorithm~\cite{Cannon2013FAR} was used to estimate the \ac{FAP}.
From the observed triggers, their distribution can be estimated for each single \ac{IFO} individually, and then used to estimate the distribution of $\rho$.  The \ac{FAP} was then calculated from this extrapolated distribution.
\end{enumerate}

All three participating algorithms were applied with both `coincidence removal' and `all samples' methods, giving six different outputs for each realisation.  In this \ac{MDC}, the three different algorithms set different lower limits on the \ac{FAP} estimation: the \ac{IFAR} algorithm adopted a lower limit of $10^{-7}$, the gstlal algorithm adopted $10^{-6}$, and the \ac{APC} algorithm set no hard lower limit.

\subsection{Estimating the error on the false alarm probability}
Essentially, all three algorithms assumed independence of the background distributions across different detectors, and in all cases estimation of the \ac{FAP} involved counting the fraction of all coincidences with larger \ac{SNR} than a given $\rho$.  The uncertainty on the calculated \ac{FAP} thus followed a simple Poisson counting error, scaling with the square root of the number of coincidences with larger \ac{SNR}.
In the limit of small \ac{FAP}, which is the most interesting region, the relative uncertainty of \ac{FAP} can be roughly approximated to be inversely proportional to the square root of the number of triggers within one \ac{IFO} that is louder than $\rho_i$. A detailed derivation and discussion of this result is available in~\cite{HamletLong}; see also~\cite{Was:2009vh} for related work on uncertainties in \ac{GW} search background estimation.

\section{Results}
In the detailed technical paper~\cite{HamletLong}, more comprehensive analysis of the \ac{MDC} results is carried out; here we focus on general conclusions. 
We examined the estimated \acp{FAP} for all $14$ experiments, applying all three algorithms with both methods of treating candidate triggers.  We then investigated our results for issues of self-consistency, precision, bias, and detection efficiency.   Signal-free experiments were exclusively adopted in subsection \ref{ppplot} while conversely only experiments containing astronomical signals were considered in subsection \ref{ROC}. All experiments were included in subsection \ref{direct} and \ref{box}.

\subsection{$p$-$p$ plots} \label{ppplot}
%
%
There were four experiments that contained no astronomical signals. 
For these  \textit{noise only} experiments, a self-consistency test can be performed by drawing a $p$-$p$ plot.  For each nominal value of \ac{FAP}, $p$, one can count the fraction within all $10^5$ realisations that has a smaller estimated \ac{FAP}.  For a self-consistent estimator on a pure noise dataset, the fraction of events having lower \ac{FAP} than $p$ should be equal to $p$, to within the counting uncertainty.
\begin{figure*}
  \vspace*{-0.3cm}
  \begin{subfigure}[b]{\columnwidth}
    \vspace*{-0.3cm}
    \includegraphics[width=0.9\columnwidth]{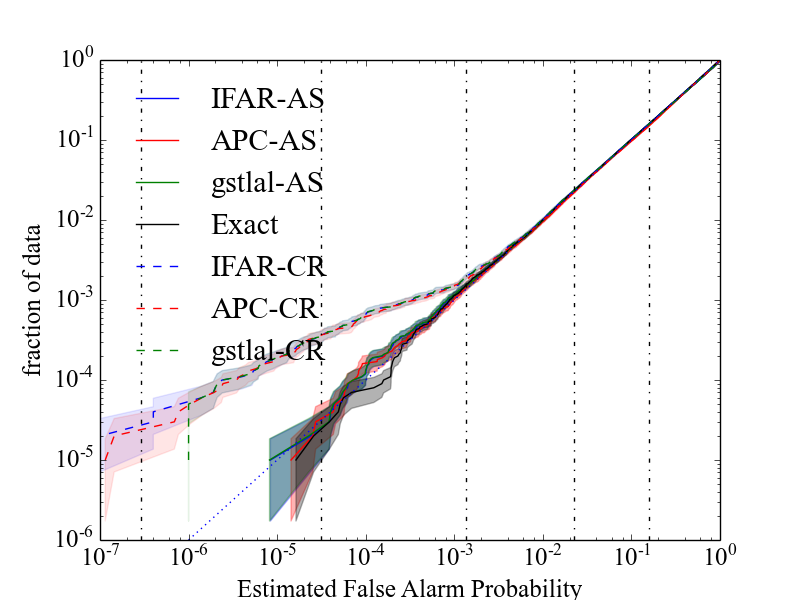}
    \caption{The $p$-$p$ plot for one experiment. The vertical dashed lines correspond to different numbers of Gaussian standard deviations from $1\,\sigma$ through $5\,\sigma$ (right to left). 
    \label{fig:12PP}}
  \end{subfigure}
  \begin{subfigure}[b]{\columnwidth}
    \includegraphics[width=1.05\columnwidth]{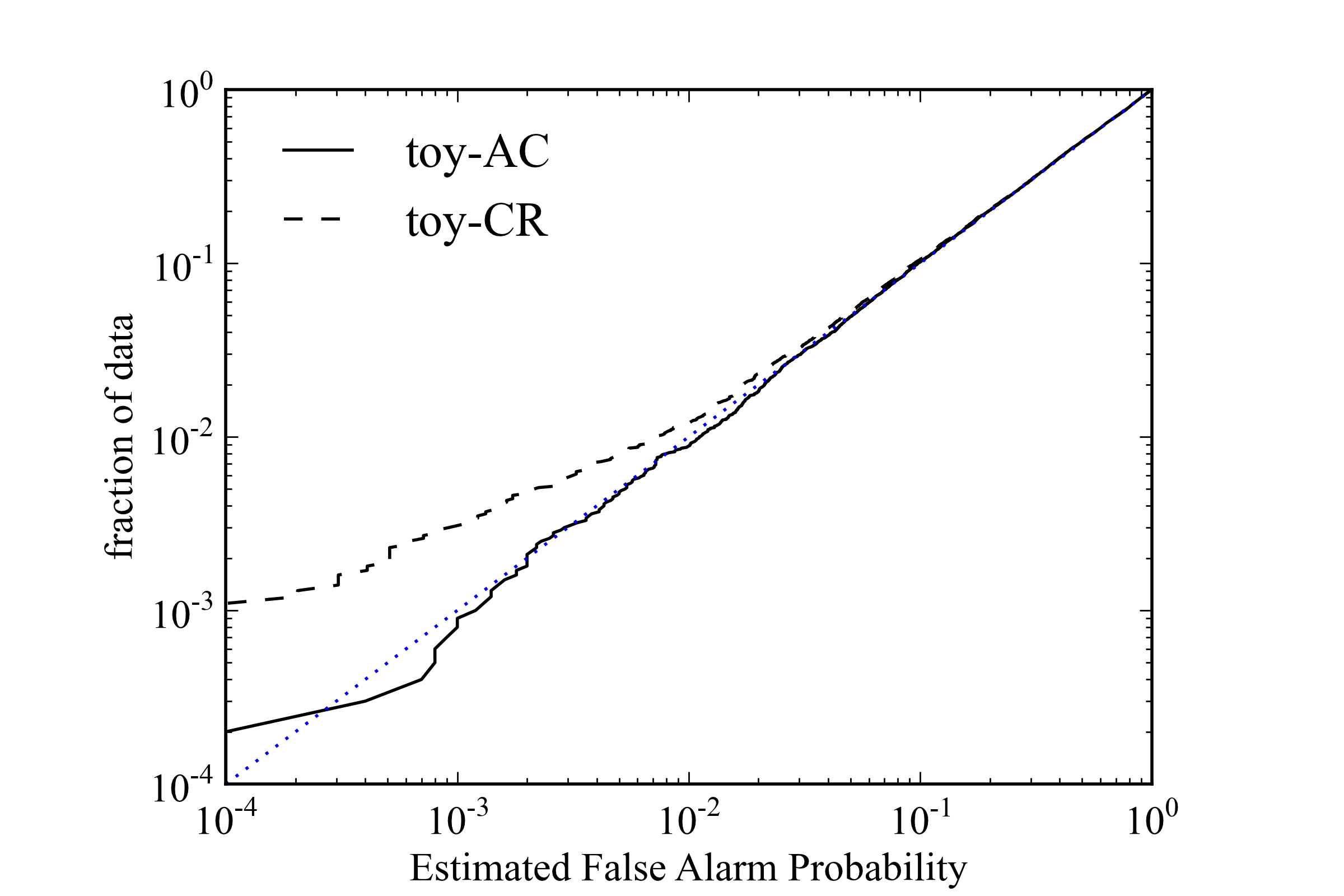}
    \caption{The $p$-$p$ plot for a toy `\ac{MDC}' in which \ac{SNR} and associated `coincidence' are assigned completely randomly. 
    \label{fig:PP_theo}}
  \vspace*{0.2cm}
  \end{subfigure}
  \caption{ The $p$-$p$ plots for one experiment with no astronomical signals (left) and for a toy model (right). The
  $x$-axis shows the estimated \ac{FAP} value, and the $y$-axis shows the fraction of events over all realisations with estimated \ac{FAP} equal to or smaller than that value.  A self-consistent estimate should lie around the diagonal line.  The solid lines represent the results for the different algorithms with the `all samples' (AS) method, while the dashed lines show results with the `coincidence removal' (CR) method. 
    \label{fig:PP}}
\end{figure*}

In Fig.~\ref{fig:12PP} we show the $p$-$p$ plot for one specific experiment where a clear discrepancy between the results for `all samples' and `coincidence removal' methods is apparent for estimated \ac{FAP} values of around $10^{-3}$ or less.  The `all samples' results show good self-consistency, while the `coincidence removal' results show a clear tendency to underestimate the \ac{FAP}, for small 
\ac{FAP} values.  This feature is universally observed, but will be most severe when the loudest single trigger forms a zero-lag coincidence and is removed -- in which case it can be understood in terms of the following heuristic explanation.  For the distributions we consider, the loudest trigger in one or other \ac{IFO} may often be a moderate outlier above the bulk of the noise distribution.  If seen in combination with this loudest trigger, almost every single trigger from the other \ac{IFO} would produce a loud coincidence. 
Thus, if the loudest trigger is removed, essentially the \ac{FAP} will be underestimated by $\mathcal{O}(1/N)$, where $N$ is the number of triggers in one \ac{IFO}.  For $n$ trials (where $n\ll N$) one can expect this scenario to affect $\sim n/N$ of the realisations. 
Taking a typical value of $n=10$ and $N=10^4$, would mean that a fraction of the loudest $10^{-3}$ realisations would be affected by this underestimation -- consistent with our results.  
Although the probability that the loudest trigger in one \ac{IFO} forms a random coincidence with a noise trigger in the other is small, it is \emph{larger} than the factor by which the \ac{FAP} could be underestimated as a consequence of its removal. 

%
%
We also investigated if this observed breakdown of self-consistency could be associated with the timing of the triggers.  We designed a simple replica of the \ac{MDC} in which only \ac{SNR} was involved, so that the `pairing' of coincident triggers was completely random.  Both `all samples' and `coincidence removal' methods were used to estimate the \ac{FAP} in this case.  Fig.~\ref{fig:PP_theo} shows an example where $N=100$ and $n=1$.  As predicted by our heuristic explanation, the discrepancy of the `coincidence removal' method becomes apparent when $n/N = 10^{-2}$.
We conclude, therefore, that the `coincidence removal' method's loss of self-consistency has its origin in the methodology itself, i.e.\ the exclusion of candidate samples which substantially affects the background estimate in a small number of cases.

%
%
\subsection{Direct comparison}\label{direct}
\begin{figure*}
  \centering
  \vspace*{-0.5cm}
  \begin{subfigure}[b]{\columnwidth}
    \centering
    \includegraphics[width=\columnwidth]{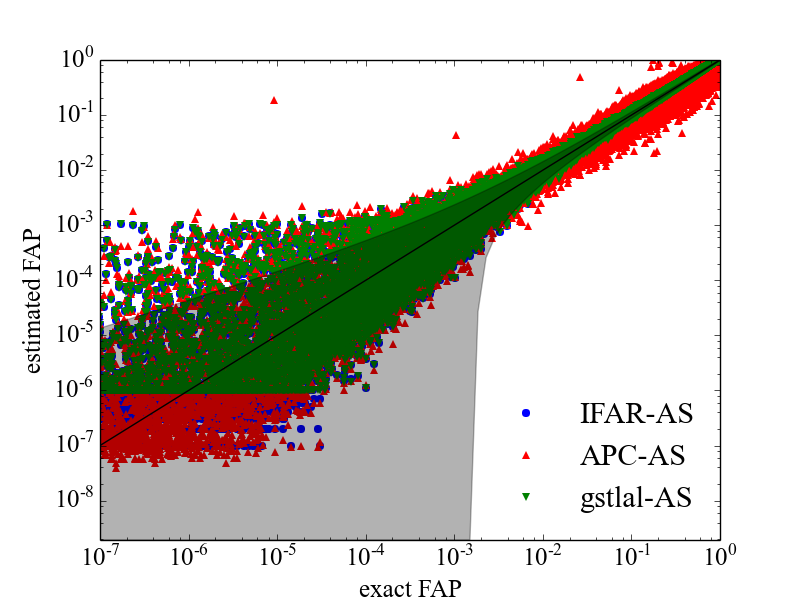}
    \caption{`All samples' comparison.}
    \label{fig:cmp9rm}
  \end{subfigure}%
  \begin{subfigure}[b]{\columnwidth}
    \includegraphics[width=\columnwidth]{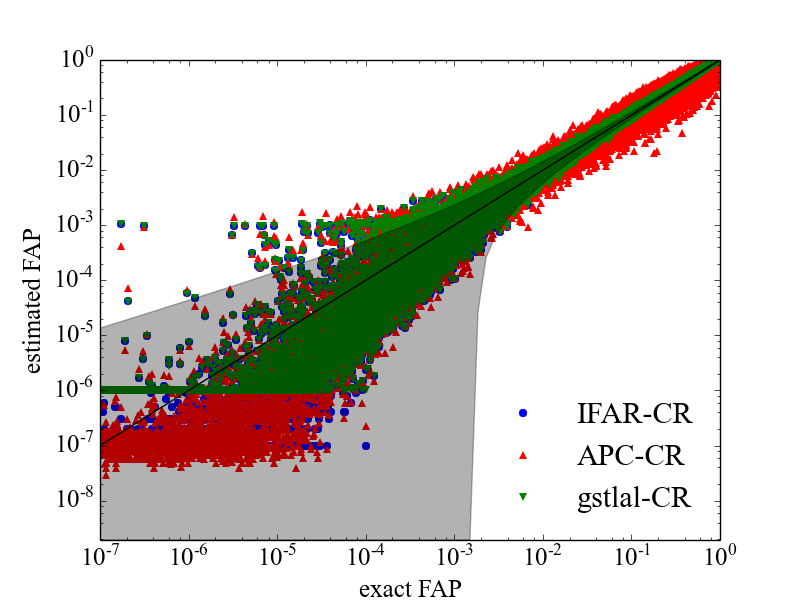}
    \caption{`Coincidence removal' comparison.}
    \label{fig:cmp9nonrm}
  \end{subfigure}
  \caption{ Directly comparing \ac{FAP} estimates with the exact \ac{FAP} for one typical experiment.
  Shaded regions represent predicted uncertainty.
  Different methods give comparable results with visible difference of lower boundary due to \emph{a priori} choice. 
  \ac{APC} method shows larger scatter in larger \ac{FAP} due to lower accuracy used, which does not affect the result of interesting small \ac{FAP} regions.
  Left panel shows results for `all samples' (AS) method and right panel for `coincidence removal'  (CR).
}
  \label{fig:cmp9}
\end{figure*}

We also compared estimated \ac{FAP} values with exact \ac{FAP} values using all experiments.  All three algorithms show similar behaviour; in each case, however, different results were obtained for the two different methods of treating candidate triggers. 
One typical result is shown in Fig.~\ref{fig:cmp9}, where the general pattern displayed is that the smaller the \ac{FAP}, the larger the relative uncertainty in its estimated value.  The spread of relative uncertainties is consistent with predictions.

It can be seen that, in the small \ac{FAP} limit, by removing all zero-lag coincidences the \ac{FAP} values are much more likely to be underestimated.  Under the condition that the exact FAP is small, but bounded by the lower limit of zero, any underestimation can only trivially decrease the expected value, while any potential overestimation would contribute disproportionately.
The conditional probability distribution $p({\rm FAP}_{\rm estimate} | {\rm FAP}_{\rm true}\ll 1, {\rm Removal})$ would thus naturally become skewed, and it is very likely that such an estimator could report an estimate of, say, $10^{-5}$, when the exact \ac{FAP} is actually $10^{-4}$. 

%
%
\subsection{Box plot}\label{box}
\begin{figure*}
  \centering
  \vspace*{-0.5cm}
  \begin{subfigure}[b]{\columnwidth}
    \centering
    \includegraphics[width=\columnwidth]{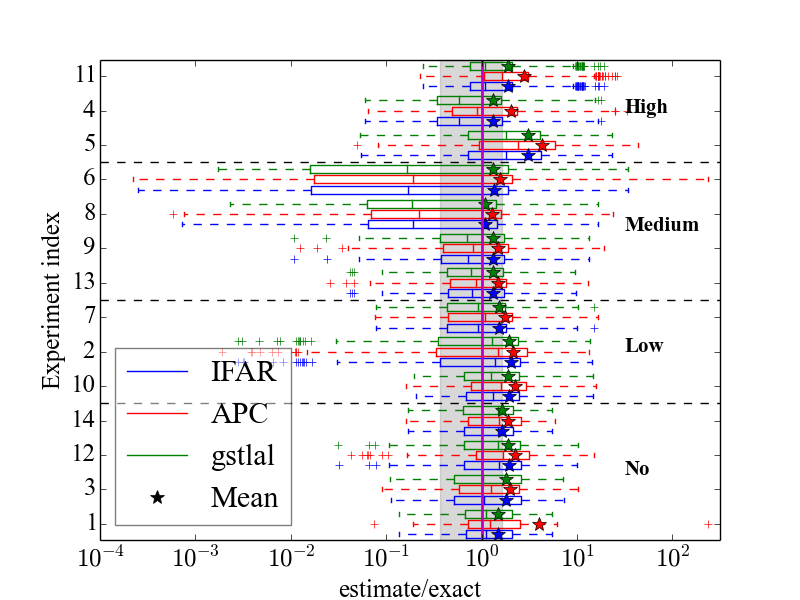}
    \caption{Box plots based on `all samples'}
    \label{fig:box-4rm}
  \end{subfigure}
  \begin{subfigure}[b]{\columnwidth}
    \centering
    \includegraphics[width=\columnwidth]{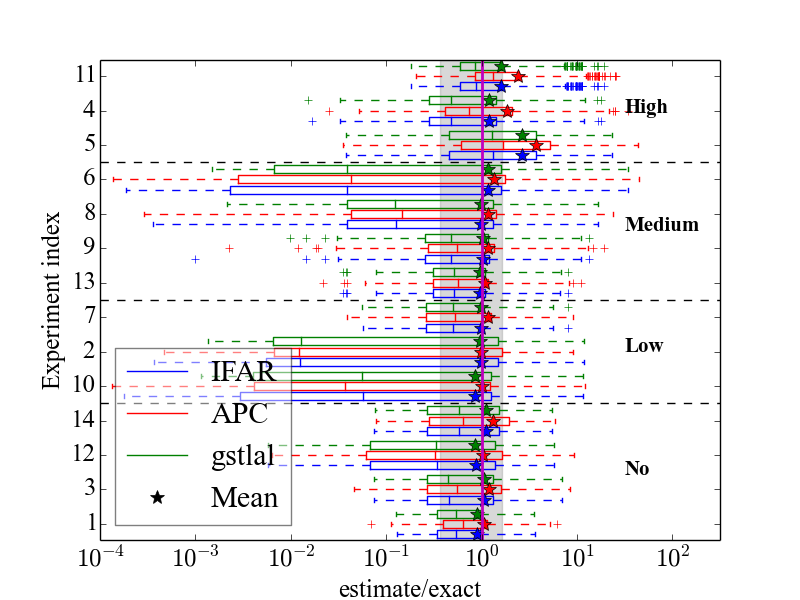}
    \caption{Box plots based on `coincidence removal'}
    \label{fig:box-4nonrm}
  \end{subfigure}
  \caption{Box plots of ratios between estimated and exact \ac{FAP} values, based on realisations with exact \acp{FAP} $\in[10^{-4},10^{-3}]$.
  The $14$ experiments are ordered by descending astrophysical signal rates. The shaded region indicates the expected uncertainty in the ratio.
  The middle box indicates the interquartile range, the thin vertical line within the box is the median, and the star represents the mean.
  }\label{fig:box-4}
\end{figure*}

In order to investigate the bias of the \ac{FAP} estimates,  using all experiments we selected realisations with exact \ac{FAP} values in the range $10^{-4}-10^{-3}$.
Fig.~\ref{fig:box-4}, shows box plots for the ratio of estimated \ac{FAP} and exact \ac{FAP}.  We can see that for the `all samples' method, for all algorithms, the median estimators are essentially unbiased within sampling uncertainty while the mean values are generally overestimated. On the other hand, results from the `coincidence removal' method show unbiased mean values of the \ac{FAP} estimator, but median values tend to be underestimated.  This result shows vividly how an unbiased estimator of \ac{FAP}, in the small \ac{FAP} limit will nonetheless lose self-consistency for the majority of realisations.

%
%
\subsection{Receiver Operating Characteristic}\label{ROC}
Finally,  using only those experiments containing astronomical signals, we plotted changed the "plotted" to "obtained" \ac{ROC} plots to compare the detection efficiency of our different algorithms using both methods of treating candidate triggers.  In these plots the \ac{FPR} is the fraction of noise realisations with \ac{FAP} smaller than a certain threshold, and \ac{TPR} is this same fraction for astronomical signals.  The closer a \ac{ROC} curve is located to the top left corner of the diagram, the higher is the detection efficiency of the estimator.

Our results showed that all algorithms gave ROC curves with similar behaviour, and for each algorithm the difference between the ROC curves for the two methods of treating candidate triggers lay mostly within the sampling uncertainty.  There were several specific experiments where the `all samples' method was found to have higher \ac{TPR} for the smallest \ac{FPR}, and the difference was larger than the numerical uncertainty. Further investigation indicated that these discrepancies between the two methods occurred at similar \ac{FAP} values as the discrepancies that arose in the $p$-$p$ plots.
By falsely assigning too much background noise with a low \ac{FAP}, the detection of signals with the `coincidence removal' method is indeed rendered more difficult.  However, this feature is not universally observed for all experiments.

\section{Summary and Conclusions}
We have set up and perfomed a \ac{MDC} to investigate the impact of removing zero-lag coincidences on the estimation of \ac{FAP} for GW candidates. 
$14$ experiments were conducted, each comprising $10^5$ realisations, covering a wide range of background distribution complexity as well as widely different astronomical signal rates.  Not all experiments were designed to represent our understanding of the signals and backgrounds expected for the current generation of ground-based GW detectors; some extreme cases were included to test the robustness of current \ac{FAP} estimation methods.  Three different algorithms for background estimation were used, each using two methods of treating candidate triggers, namely `coincidence removal' and `all samples'.  Throughout all experiments, the three different algorithms showed substantially similar behaviour; obvious differences arose only due to different choices of \ac{FAP} cutoff and calculation accuracy.  

However, the two different methods of treating candidate triggers showed clear differences in the estimated \ac{FAP} values -- with the `coincidence removal' method generally having an unbiased mean value while the `all samples' method was self-consistent.
We demonstrated that, for the most interesting regime of small \ac{FAP} values, self-consistency and unbiasedness for the mean \textit{cannot} be achieved simultaneously.  

We recommend that, for the first detections of \ac{GW} from a previously unobserved source, particularly when signal rates are highly uncertain, a strict threshold should be imposed on the self-consistent \ac{FAP} calculated via the `all samples' method.  
Conversely, the mean-unbiased nature of the `coincidence removal' method means its estimate of the \ac{FAP} as a function of the search ranking statistic is more informative about the background distribution.  

With the expected increase in sensitivity of the Advanced detector network~\cite{Aasi:2013wya}, the rate of true signals is expected to increase over time.
With a higher rate of astronomical events, {\it a priori} the loudest candidate event is correspondingly more likely to be signal than noise. Therefore, in the longer run, we would expect to place more importance on mean unbiasedness than on self-consistency, which favours the `coincidence removal’ method in searches where the rate of detectable signals is known to be high.
However, for cases where $>1$ loud signals occur per experiment, we anticipate that Bayesian methods based on modelling signal and noise distributions simultaneously, with a well-defined signal prior~\cite{Farr:2015} will be more suitable.
%

We also computed a heuristic estimate of the relative uncertainty in the \ac{FAP} value: the square root of the number of triggers within one \ac{IFO} that are louder than $\rho_i$.  This  predicted uncertainty is roughly consistent with the numerical scatter observed in our realisations.  Moreover, the deviation from self-consistency affects the \ac{FAP} especially strongly for values less than or equal to $n/N$, where $n$ is the expected number of coincidences and $N$ is the total number of triggers in one \ac{IFO}.  We note that, due to the expedient choice of parameters in our simulations, our quantitative numerical conclusions are strictly only valid in the context of the \ac{MDC}, although they do offer an instructive indication of the expected behaviour for more realistic cases.

\begin{acknowledgments}
The authors have benefited from discussions with many \ac{LSC} and Virgo 
collaboration members and ex-members including Drew Keppel and Matthew West. We acknowledge the use of LIGO Data Grid computing facilities for the generation and analysis of the \ac{MDC} data.
C.~M.\ is supported by a Glasgow University Lord
Kelvin Adam Smith Fellowship and the Science and Technology Research Council (STFC) grant No. ST/ L000946/1. M.~H. is supported by STFC 
grant No. ST/L000946/1. J.~V.\ is supported by STFC grant No. ST/K005014/1.
\end{acknowledgments}

\bibliography{../masterbib,../cbc-group}
\end{document}